\documentclass[%
times,
8pt,
5p,
amsmath,amssymb,
reprint,%
aip,
lualatex -synctex=1 -interaction=nonstopmode --shell-escape 
]{article}

\usepackage{arxiv}

\usepackage[utf8]{inputenc} 
\usepackage[T1]{fontenc}    
\usepackage[hidelinks]{hyperref}       
\usepackage{url}            
\usepackage{booktabs}       
\usepackage{amsfonts}       
\usepackage{nicefrac}       
\usepackage{microtype}      
\usepackage{lipsum}		
\usepackage{graphicx}
\usepackage{doi}

\usepackage{pdfpages}

\usepackage[round, sort, numbers, compress]{natbib}
\setcitestyle{square}
\title{Fundamentals of Scanning Surface Structuring by Ultrashort Laser Pulses: From Electron Diffusion to Final Morphology}
\date{\today}

\author{ \href{https://orcid.org/0000-0001-5051-7209}{\includegraphics[scale=0.06]{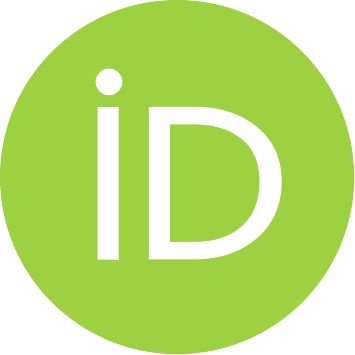}\hspace{1mm}F. Nyenhuis}\thanks{fabian.nyenhuis@gmail.com} \\
	Robert Bosch GmbH, Bosch Research,\\
	Postbox 30 02 40, Stuttgart, 70442, Germany \\
	\And
	\href{https://orcid.org/0000-0002-6616-7351}{\includegraphics[scale=0.06]{Figures/orcid.pdf}\hspace{1mm}P. N. Terekhin} \\
	Department of Physics and Research Center OPTIMAS,\\
	Technische Universität Kaiserslautern,\\
	Erwin-Schrödinger-Str. 46, Kaiserslautern, 67663, Germany \\
	\AND
	\href{https://orcid.org/0000-0001-5463-914X}{\includegraphics[scale=0.06]{Figures/orcid.pdf}\hspace{1mm}T. Menold}\\
	Robert Bosch GmbH, Bosch Research,\\
	Postbox 30 02 40, Stuttgart, 70442, Germany \\
	\And
	B. Rethfeld \\
	Department of Physics and Research Center OPTIMAS,\\
	Technische Universität Kaiserslautern,\\
	Erwin-Schrödinger-Str. 46, Kaiserslautern, 67663, Germany \\
	\And
	A. Michalowski\\
	Robert Bosch GmbH, Bosch Research,\\
	Postbox 30 02 40, Stuttgart, 70442, Germany \\
	\And
	J. L'huillier \\
	Photonik-Zentrum Kaiserslautern e.V. and Research\\
	Center OPTIMAS, Technische Universität Kaiserslautern,\\
	Kohlenhof~Strasse~10, Kaiserslautern, 67633, Germany \\
}

\renewcommand{\headeright}{ }
\renewcommand{\shorttitle}{\textit{}}

\hypersetup{
pdftitle={Fundamentals of Scanning Surface Structuring by Ultrashort Laser Pulses: From Electron Diffusion to Final Morphology},
pdfsubject={ },
pdfauthor={F. Nyenhuis, P.N. Terekhin, T. Menold, B. Rethfeld, A. Michalowski, L'huillier },
pdfkeywords={Ultrashort pulse milling, Laser-induced periodic surface structures, Surface bumps, Heat accumulation, Two-temperature model, Surface plasmon polariton},
}

\begin{document}
\maketitle

\begin{abstract}
Industrial use of ultrashort-pulse surface structuring would significantly increase by an effective utilization of the average laser powers available currently. However, the unexplained degradation of surfaces processed with numerous pulses at high average laser power makes this difficult. Based on a systematic experimental study, the structure formation underlying such surface degradation was investigated. We presented a hierarchical structural formation model that bridges the gap between laser-induced periodic surface structures and surface degradation. Contrary to expectations based on previous research, we observed less structure formation on titanium for high laser fluences. As a possible reason, enhanced electron diffusion with increasing intensity was investigated within the framework of the two-temperature model. Our findings provide a deeper understanding of the microscopic mechanisms involved in surface structuring with ultrashort pulses.
\end{abstract}
%

\section{\label{sec:Introduction}Introduction}
Surface structuring using ultrashort laser pulses has been intensively investigated since the early nineties \cite{chichkov1996femtosecond, nolte1997ablation, preuss1995sub, momma1997precise,chong2010laser}. Since then, ultrashort pulse (usp) lasers have been considered a high-precision, flexible,  and easily automated tool for industrial material processing. However, the use of usp lasers as a manufacturing tool is often not economical owing  to their limited ablation rates. In particular, in multi-pulse surface structuring of industrially relevant metals such as titanium, iron, or their alloys, the maximum ablation rates are limited to a few cubic millimeters per minute \cite{neuenschwander2012optimization,vzemaitis2019rapid,finger2015high,vzemaitis2018advanced}. Consequently, usp milling cannot meet the cost requirements of industrial series production for the majority of applications.

An unexplained degradation of surfaces processed at high average power \cite{bauer2015heat,vzemaitis2019rapid,fraggelakis2017texturing} results in limited ablation rates. This degradation is manifested by an increased growth of surface structures, which ultimately leads to bumpy surfaces. The occurrence of bumps was associated with exceeding the material-specific critical saturation temperature $T_{\mathrm{c}}$ that  results from the heat accumulation of successively applied laser pulses \cite{bauer2015heat}. Thus, a critical spatio-temporal pulse overlap was defined to prevent an excess of $T_{\mathrm {c}}$ \cite{bauer2015heat}. 

Given that  the dynamics of scanners are limited by finite acceleration, or more precisely by the jerk, a critical spatio-temporal pulse overlap ultimately leads to limited ablation rates \cite{jaeggi2016time}. Therefore, there is a great interest in industry to understand the mechanisms of structure formation causing bumpy surfaces. This remains an open research topic. Although some previous studies mention the connection of bump occurrence with heat accumulation \cite{fraggelakis2017texturing,faas2018prediction,bauer2015heat,ackerl2020evolution}, there is still no conclusive explanation for the process underlying the formation of bumps.

Therefore, the aim of this study  was to identify potential formation mechanisms of bumps. For this purpose, we conducted ablation experiments on titanium (Ti) that revealed a relation between the empirical temperature $T_{\mathrm{c}}$ and the convective growth of a laser-induced periodic surface structure (LIPSS)  type called grooves. Using analytical calculations and microstructure analysis, we aimed to understand the origin of the grooves, as they are the key structure for the transition from nanoscale LIPSS to microscale bumps. 

The paper is organized as follows. In Section 2, a \textit{hierarchical structure formation model} is presented to provide an overview of the whole bump formation process. Within this model, the hierarchy, i.e., the predominant type of surface structures, changes as a function of the number of pulses and the spatio-temporal pulse overlap. This is followed by related experiments and material analyses in Sections 3.1 to 3.3. Section 3.4 presents experiments on reduced structure formation at higher fluences or intensities. Within the framework of the two-temperature model, in Section 3.5 we discuss potential reasons for reduced structure formation at higher intensities. The simulation results indicate a potentially strong influence of increasing intensity on the transient change of optical properties, suggesting an electromagnetic origin of the structures. Our findings have extensive consequences for the scaling of ablation rates and development of alternative process strategies~\cite{nyenhuisdual,nyenhuis2020surface}.
\section{\label{sec:Theo_view}Hierarchical structure formation model}

\begin{figure}[h!]
    \centering
    \includegraphics{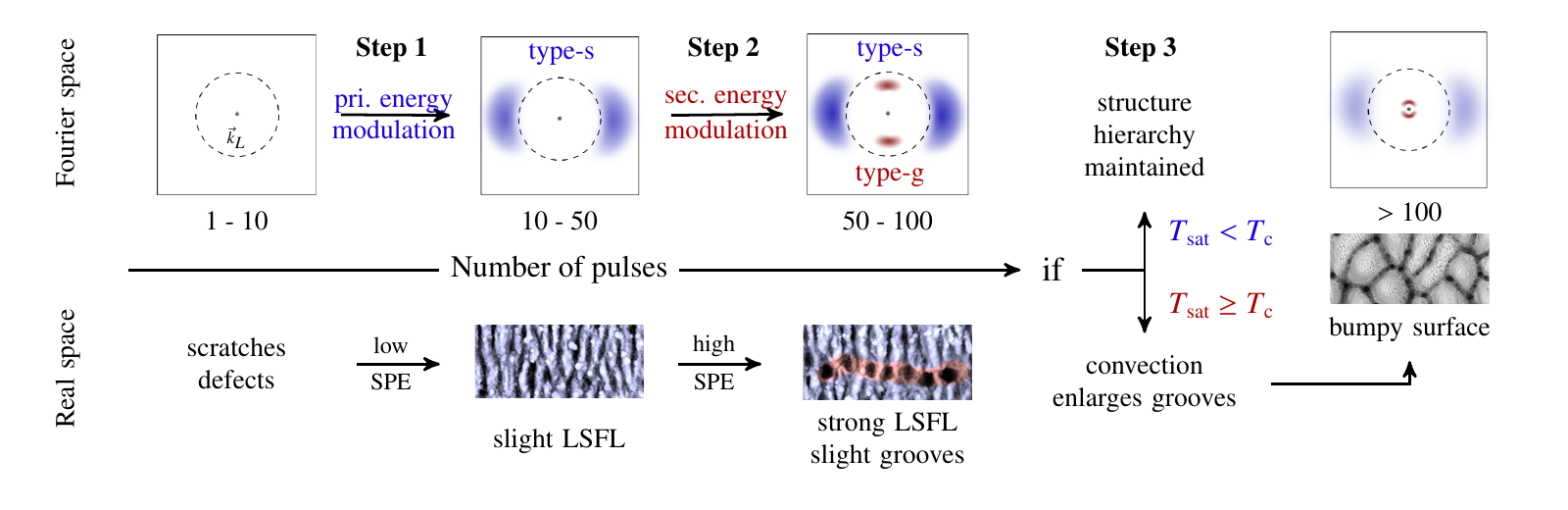}
    \vspace{-12pt}
    \caption{Hierarchical structure model in usp laser milling as a function of the number of applied pulses. The top row shows characteristic signals of different structure types in the Fourier space, with the dashed circle representing the wave vector of the laser light. The type-s signal (scattering, manifested as LSFL-I) is shown in blue and the type-g signal (grooves, supra-wavelength-sized) in red (notation as in \cite{zhang2020laser,skolski2013inhomogeneous}). The bottom row shows SEM images of the corresponding structure types that indicate some causes involved in the formation process (SPE stands for surface plasmon excitation). The SPE and near-surface scattering of light leads to a modulation of the deposited energy, denoted as \textit{primary} and \textit{secondary energy modulation} in the Fourier space. With increasing number of pulses, the dominant structure changes from LSFL to grooves, and ultimately, if the saturation temperature within the interaction zone exceeds a critical value $T_{\mathrm{sat}}$ $ \geq$ $T_{\mathrm{c}}$, from grooves to bumps.\label{fig:sketch}}
\end{figure}

A \textit{hierarchical structure formation model} is introduced in Figure \ref{fig:sketch} to explain the steps involved in the bump formation process at moderate fluences and high repetition rates. The individual hierarchy levels are organized according to the number of applied pulses. Each surface structure has a characteristic signal in the Fourier space (top row in Figure~\ref{fig:sketch}) according to its period and orientation. The corresponding surface structure is shown in a scanning electron microscopy (SEM) image in the bottom row (real space) in Figure~\ref{fig:sketch}. The steps  are discussed in more detail below.

\subsection{Step 1: low-spatial-frequency structure formation}

Starting from a smooth surface, low-spatial-frequency LIPSS (LSFL) with a period $\Lambda_\mathrm{s}$ close to the laser wavelength $\lambda_{\mathrm{L}}$ gradually form in a scanning multi-pulse ablation process due to an inhomogeneous energy deposition (denoted as \textit{primary energy modulation} in Figure \ref{fig:sketch}). The shape of this inhomogeneous energy deposition (IED) is caused by a modulation of the intensity distribution near the surface. For metals, this modulation results mainly from an interference between the incident light and surface scattered light assisted by surface plasmon polariton (SPP) excitation \cite{rudenko2019self,lalanne2009microscopic,nikitin2010surface,zhang2015coherence,rudenko2019light,skolski2014modeling,he2016generation,bonse2016laser,sipe1983laser,benhayoun2021theory,ackerl2020observation,ng2019plasmonic,wang2021high,plech2009femtosecond,oltmanns2021Influence,terekhin2022key}. In principle, the IED can be partly calculated with analytical expressions \cite{sipe1983laser,nikitin2010surface,terekhin2020influence} or with numerical approaches such as finite-difference time-domain (FDTD) simulations  \cite{lalanne2009microscopic,skolski2014modeling,zhang2015coherence,rudenko2020high,rudenko2019light}. 

A change in surface topography as machining progresses is expected to affect the IED \cite{ng2019plasmonic,wang2021high,zhang2015coherence}; this change is referred to as inter-pulse feedback \cite{zhang2015coherence,zhang2020laser,skolski2014modeling}. High-spatial-frequency LIPSS (HSFL) or high-frequency modulations on a scale $\ll$~$\lambda_{\mathrm{L}}$ are assumed to play a minor role in comparatively macroscopic laser milling. Therefore, for a general discussion of LIPSS, the reader is referred to review articles available in the literature \cite{bonse2020maxwell,stoian2020advances,bonse2016laser,vorobyev2013direct}. 

\subsection{Step 2: groove formation from LSFL}

With a further increase in the number of pulses ($N_\mathrm{p} \geq$~50), slight grooves grow from the LSFL. These grooves cause a type-g signal in the Fourier space, colored in red in Figure \ref{fig:sketch} and perpendicular to the blue type-s signal of the LSFL. A groove is also colored in red in the corresponding SEM image.

Although grooves were already mentioned in early studies on LIPSS \cite{bonse2005structure}, they are probably the least studied structures \cite{nivas2021generation}. Consequently, the origin of grooves is not yet fully understood, and only recently several articles have been published on this issue \cite{he2016generation,nivas2021generation,tsibidis2016convection,fraggelakis2017texturing,nyenhuis2021ultrashort}. Existing modeling approaches may be roughly divided into two groups: those that use hydrodynamic effects such as convection rolls without additional IED besides the contributions for LSFL formation \cite{tsibidis2015ripples,tsibidis2016convection,tsibidis2018modelling}, and those that use only IED \cite{skolski2014modeling,zhang2015coherence,he2016generation,zhang2020laser} or an interaction in addition to hydrodynamic effects \cite{rudenko2020high}. Interestingly, grooves already appeared in a study conducted by Skolski $et$  $al.$ \cite{skolski2014modeling,skolski2012laser}, who solved only Maxwell's equations using FDTD simulations. Recent studies also pointed to an electromagnetic nature of grooves \cite{rudenko2020high,zhang2015coherence,zhang2020laser}. In this context, many different electrodynamic theories may be discussed \cite{lalanne2009microscopic,rudenko2019light,rudenko2019self,nikitin2010surface,zhang2020laser}. However, the regularity of the grooves and their dependence on the laser wavelength \cite{nivas2021generation,tsibidis2018modelling} imply a direct connection to the LSFL. In general, the LSFL period $\Lambda_\mathrm{s}$ decreases and their depth increases as the pulse number increases \cite{bonse2013sub,bonse2005phase,huang2009mechanisms}. This affects SPP excitation, surface-wave propagation, and near-surface light scattering \cite{lalanne2009microscopic,huang2009mechanisms, zhang2015coherence,stankevivcius2021direct,zhang2020laser}, indicating a change in the IED as machining progresses, which may lead to a new \textit{secondary modulation} (see Figure \ref{fig:sketch}). 

\subsection{Step 3: convection-driven growth of grooves}

The third step includes a condition (see Figure \ref{fig:sketch}) for the further growth of grooves: if the saturation temperature $T_{\mathrm{sat}}$ resulting from the heat accumulation of successively applied pulses is high ($T_{\mathrm {sat}} \geq T_{\mathrm{c}}$), the grooves grow owing to convection mechanisms; this leads ultimately to the excavation of bumps. At lower heat accumulation ($T_{\mathrm {sat}} < T_{\mathrm{c}}$), faster solidification hinders convection, causing the structural hierarchy to end at shallow grooves. Increased ablation caused by reflections within the grooves may also contribute to their further growth and finally to the shape of the bumps \cite{qin2012comparison}.

The accumulation of defects at higher fluences or at levels below the ablation threshold may also lead to surface degradation \cite{fraggelakis2017texturing,villerius2019ultrashort}. Because such effects are not related to the critical spatio-temporal pulse overlap, they do not lie within the scope of this article.

\section{\label{sec:Experiments}Results and Discussion}

Next, experimental results are presented along with the steps of the \textit{hierarchical structure model}. The groove formation, rarely examined to date, was investigated at MHz repetition rates as a function of the spatio-temporal  pulse overlap and pulse number. Materials and methods are described in the supplementary material.

\subsection{\label{ssec:LIPPS in Laser milling}From LSFL to grooves: heat-accumulation-related structure formation}

In our experiments, the critical parameter, namely the spatio-temporal pulse overlap, was varied by changing the scan speed $v_\mathrm{s}$. The number of scans $N$ was adjusted accordingly to keep the number of pulses $N_\mathrm{p}$ constant. Linear polarization was used, thereby enabling easy distinction of the different structure types. 

\begin{figure}[h!]
	\centering
	\includegraphics{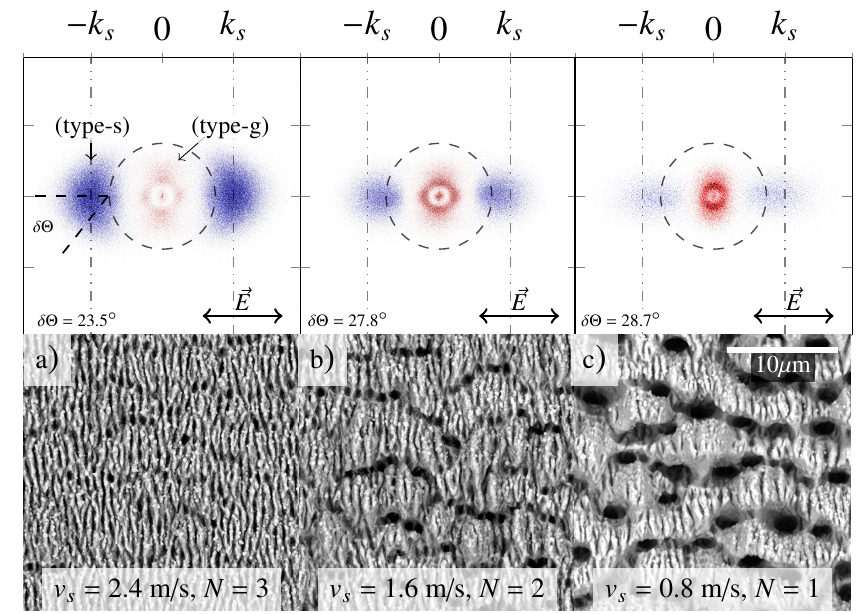}
	\caption{\label{fig:incr_vmin_ti} SEM images of milled titanium with a peak fluence $\Phi$~=~0.9~J/cm$^2$, repetition rate $f_{\mathrm{R}}$~=~1~MHz, pulse duration $\tau$~=~400~fs, spot size 2$\omega_0$~=~40~$\mu$m, hatch distance $d_h$ = 5~$\mu$m, and linear polarization . The number of scans $N$ was adjusted to $v_S$ to keep $N_\mathrm{p}$ unchanged. The dashed circle marks the wave vector of the laser wavelength $k_L$~=~2$\pi/\lambda_{\mathrm{L}}$, and the ticks show the wave vectors $k_\mathrm{s}$~=~2$\pi$/(3/4$\cdot\lambda_{\mathrm{L}}$) that should correspond to the LSFL (type-s, in blue) at the applied number of pulses \cite{bonse2013sub}. The groove features are much more noticeable  in both the FFT picture (type-g, in red) and SEM images for smaller scan speeds.}
\end{figure}

Figure \ref{fig:incr_vmin_ti} shows SEM images of the milled Ti surfaces. Above each SEM image in Figure \ref{fig:incr_vmin_ti}, the corresponding fast Fourier transformation (FFT) is shown. Although $N_\mathrm{p}$ was equal in all experiments, the surface morphology varied strongly with the scan speed. 

Figure \ref{fig:incr_vmin_ti} a) clearly shows that ripples (LSFL) appear perpendicular to the laser polarization with a corresponding wavelength $\Lambda_\mathrm{s}$ of approximately 3/4 the laser wavelength $\lambda_\mathrm{L}$~=~1030~nm. In the FFT, this surface feature is strongly expressed as a type-s signal (in blue), indicating that LSFL are the dominant structure in this case. Additionally, slight grooves run perpendicular to the ripples. The grooves generate a type-g signal, colored in red, within the dashed circle of the FFT.

In Figure \ref{fig:incr_vmin_ti} b) \& c), the grooves begin to connect. Also, the type-g features in the FFT shift to smaller wave vectors (near the center) for lower scan speeds and become more dominant. This indicates that the grooves gradually become the dominant surface structure with decreasing scan speed, and that their spatial extent continues to expand.

The type-s signal of the LSFL reduces with decreasing scan speed because the grooves encompass an increasingly larger area of the surface. In addition, the dispersion LIPSS orientation angle (DLOA) $\delta \Theta$ increases slightly with decreasing scan speed, meaning that the remaining LSFL are less parallel.

The groove formation is clearly related to the heat accumulation caused by a variation in scan speed. To investigate the dependence on $N_\mathrm{p}$, the scan speed was kept constant and $N_\mathrm{p}$ was successively increased in the following experiments. Furthermore, circular polarization was applied.

\subsection{From grooves to bumps: pulse-number-related structure formation}

\begin{figure}[t!]
	\centering
	\includegraphics{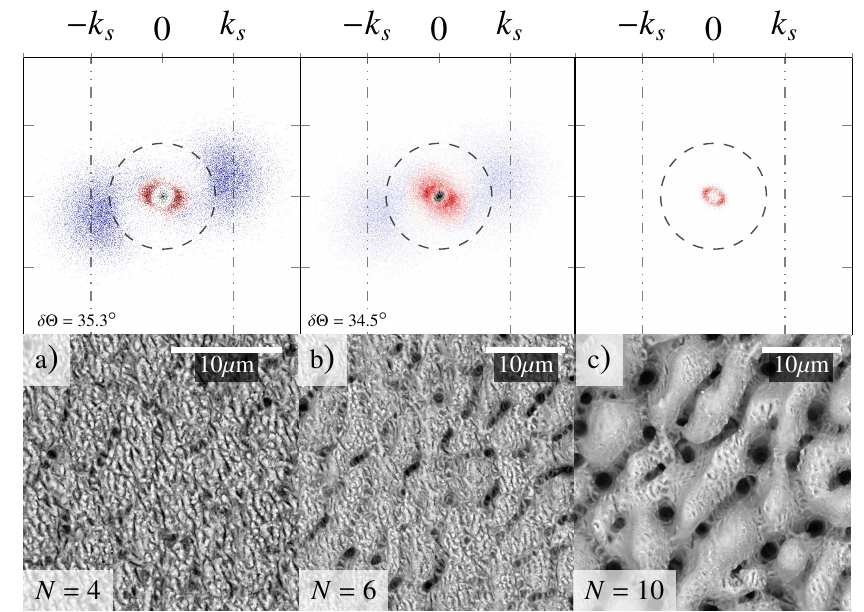}
	\caption{\label{fig:Titan_1MHz_zirkular_1mps} SEM images of milled titanium for increasing number of scans $N$, $f_{\mathrm{R}}$ = 1MHz, $\Phi$ = 0.7 J/cm$^2$, $\tau$ = 1 ps, $v_\mathrm{s}$ = 1 m/s, $d_h$ = 7 $\mu$m, and bidirectional 90$^\circ$-cross-hatch scanning with circular polarization. The number of scans $N$ is indicated in the images. The type-s (blue) and type-g signals (red) are not oriented perpendicular to each other. As $N$ increases, the grooves become more pronounced, which is also indicated in the FFTs by an increase in the type-g signal. The deepening of the grooves in c) creates a topography in which incipient bump formation can already be distinguished. }
\end{figure}

In the experiments performed, the number of pulses $N_{\mathrm{p}}$ was varied by increasing the number of scans $N$. 
The resulting surface in Figure \ref{fig:Titan_1MHz_zirkular_1mps} a) shows less regular LSFL, but this time over a larger angle $\delta \Theta$. 
Although the features of the grooves are already clearly visible in the FFT of Figure \ref{fig:Titan_1MHz_zirkular_1mps} a), they can be observed with difficulty and as slight shadows when looking at the surface. This changes as the number of scans increases, and they are already clearly visible in Figure \ref{fig:Titan_1MHz_zirkular_1mps} b). With increasing scans, the type-s signal in the FFTs decreases, and the type-g increases and shifts to smaller wave vectors. 

Despite the circular polarization, both type-s and type-g signals exhibit a preferred orientation. It is known from previous studies that the regularity of LSFL depends on the polarization orientation relative to the scanning direction \cite{gnilitskyi2017high}. Experimental results are presented in Figure~\ref{fig:Titan_1MHz_zirkular_1mps} for a 90$^{\circ}$-cross-hatch with right-left and bottom-top scan directions. There is also a preferred direction for the type-g signal that is not perpendicular to the type-s signal. The signals exhibit a relative angle of approximately 38$^{\circ}$. The non-perpendicular arrangement of type-s and type-g signals may result from an asymmetry of the intensity distribution caused by scattering processes, given that these processes  are angle- and polarization-dependent \cite{zhang2020laser,plech2009femtosecond}. 

The longer grooves for linear polarization (Figure \ref{fig:incr_vmin_ti}) indicate a correlation with the longer and stronger pronounced LSFL. Despite the lower pulse number and higher scan speed and, consequently, lower saturation temperature, the structures are more pronounced with linear polarization. According to FDTD simulations, the type-g feature in the FFT of the IED occurs only when scattering objects (e.g., LSFL) with depth $\geq$~100~nm are present \cite{zhang2015coherence}. As the depth of ripples or height of nanobumps (LSFL at circular polarization) increases, their edges become steeper, leading to a local increase in the angle of incidence. Consequently, the scattering may increase locally, making its contribution to the type-g signal likely. However, further investigation is necessary to identify the electromagnetic origin of the grooves.

For a greater number of scans (Figure \ref{fig:Titan_1MHz_zirkular_1mps} b \& c), depressions become visible at the ends of the grooves. Some areas of the SEM image in Figure~\ref{fig:Titan_1MHz_zirkular_1mps}~c) already appear to be bumps \cite{bauer2015heat}, which would reduce the ablation rate \cite{nyenhuisdual}. In the corresponding FFTs, the same orientation of the type-g signal as before is visible. Consequently, the orientation of the type-g signal is maintained for different numbers of scans, and shifts successively to smaller wavenumbers. As soon as the grooves become significantly larger than the laser wavelength $\Lambda_{\mathrm{G}} \gg \lambda_{\mathrm{L}}$, further evolution may be considered via polarization-dependent reflection behavior \cite{qin2012comparison}. 

\begin{figure}[t!]
	\centering
	\includegraphics{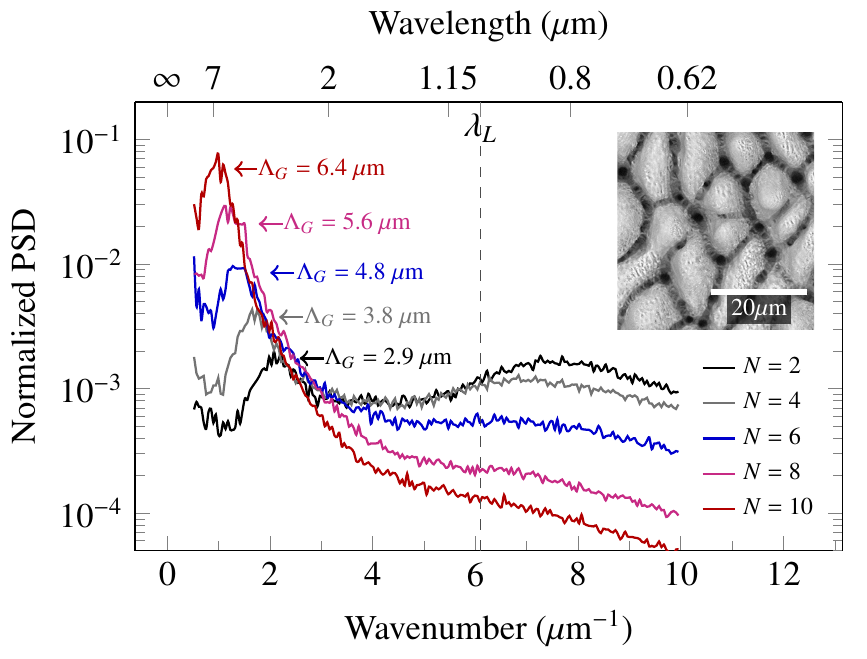}
	\caption{\label{fig:PSD_TITAN} Normalized radial PSD of milled titanium surfaces for an increasing number of scans $N$. The process parameters are shown in Figure \ref{fig:Titan_1MHz_zirkular_1mps} . For small values of $N$, the part of the spectrum to the right of the laser wavelength $\lambda_{\mathrm{L}}$ (dashed help line) associated with the LSFL dominates. As $N$ increases, the LSFL part decreases and the peak associated with the grooves becomes dominant and shifts to smaller wavenumbers. The inset shows an SEM image of the surface after 20 scans.}
\end{figure}

Given that the shift of the signals cannot be easily inferred from  the two-dimensional FFT, the normalized radial power spectral density (PSD) is additionally plotted in Figure~\ref{fig:PSD_TITAN}. An increase in the PSD signal for smaller wavenumbers is observed as the number of scans increases. In addition, the high wavenumber signal around $k_\mathrm{s}$~=~7.5~$\mu$m$^{-1}$ ($\Lambda_\mathrm{s}$~$\approx$~850~nm) associated with the LSFL decreases. This illustrates well that as the number of scans increases, grooves replace LFSL as the dominant surface structure. Furthermore, the peak associated with the grooves shifts to smaller wavenumbers (values are shown in the plot). 
As the grooves progressively deepen and broaden, their preferred direction is maintained, resulting in the formation of elliptical-shaped bumps. The elliptical shape of this final type of surface structure is shown as an inset in Figure \ref{fig:PSD_TITAN}. A polarization-dependent shape development of surface structures, e.g., ellipsoidal conical spikes, was already reported in previous studies \cite{zhu2006effect,ahmmed2015introducing}. 

On a surface strongly covered with grooves, inhomogeneous ablation may be favored. On the one hand, there are areas where the projection of the initial fluence distribution onto slopes reduces the effective fluence, and on the other hand, reflections locally cause higher fluences \cite{qin2012comparison}. This results in higher fluence within the grooves that digs the bumps out.  

Consequently, the convective contribution to the overall process of bump formation would be mainly associated with the groove formation. To evaluate this hypothesis and investigate whether the bumps rather than the grooves may have been formed by convection, focused ion beam (FIB) cuts were made for a surface covered with bumps.

\subsection{\label{ssec: Bumps on Titanium} Convective contributions to bump formation}

\begin{figure}[t!]
	\centering
	\includegraphics{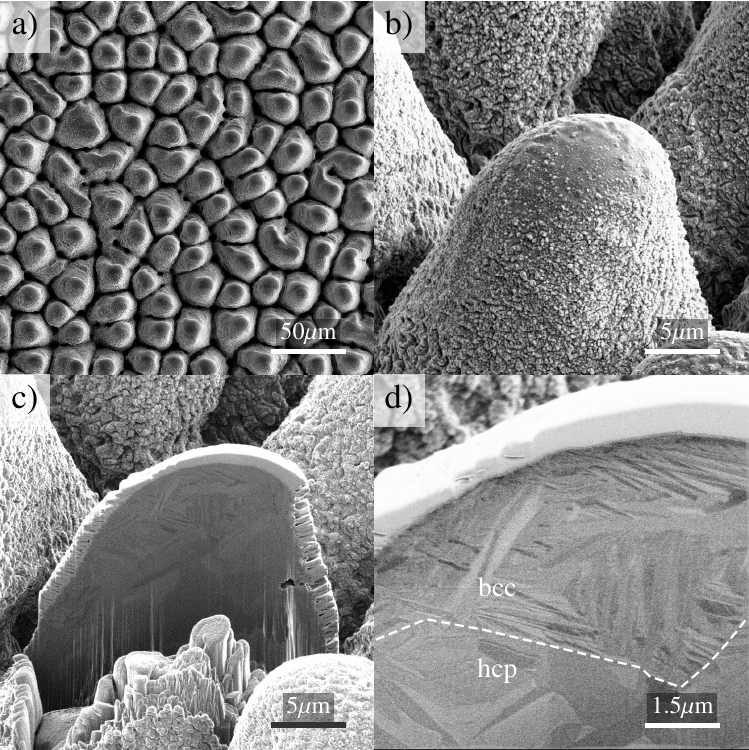}
	\caption{\label{fig:FIB_Bump} FIB  cut of a representative bump on titanium after laser milling with circular polarization, $f_{\mathrm{R}}$ = 814 kHz, $\tau$ = 3.3 ps, $\Phi$~=~1.3~J/cm$^2$, $d_h$ = 10$\mu$m, $v_\mathrm{s}$ = 1 m/s, and 40 scans. Picture a) shows a top-view SEM image of the machined surface in low magnification. Picture b) shows an SEM image of the bump selected for the FIB cut. To create an intense grain orientation contrast, an ion beam of 30 kV was used for imaging in pictures c) \& d). The dashed white line in picture~d) indicates the boundary between the hcp and bcc  systems. The white top layer is a deposited platinum coating.}
	\vspace{-4.25mm}
\end{figure}

Figure \ref{fig:FIB_Bump} shows the top view of a bumpy surface and  magnified image of a bump selected for the FIB cut. Prior to the cut, the whole surface was covered with a platinum protection layer, which is visible as a white layer in Figure \ref{fig:FIB_Bump} c) \& d). To create an intense grain orientation contrast, an ion beam was used for imaging in Figure \ref{fig:FIB_Bump} c) \& d). The upper grains exhibit a needle-like shape. Although these grains indicate a heat-affected zone, no clear melt pool contour is visible. Therefore, it is not possible to distinguish whether the region was eventually melted or whether the phase boundary between hexagonal close-packed (hcp) and body-centered crystal (bcc) systems of Ti ($\beta$ transus  temperature $\geq$ 882$^\circ$ C \cite{leyens2006titanium}) was merely crossed. Hence, additional energy-filtered transmission electron microscopy (EF-TEM) images were taken for the samples. An amorphous carbon protection layer was applied prior to the EF-TEM analysis. The protocol is further described in the supplementary material. 

\begin{figure}[t!]
\centering
	\includegraphics{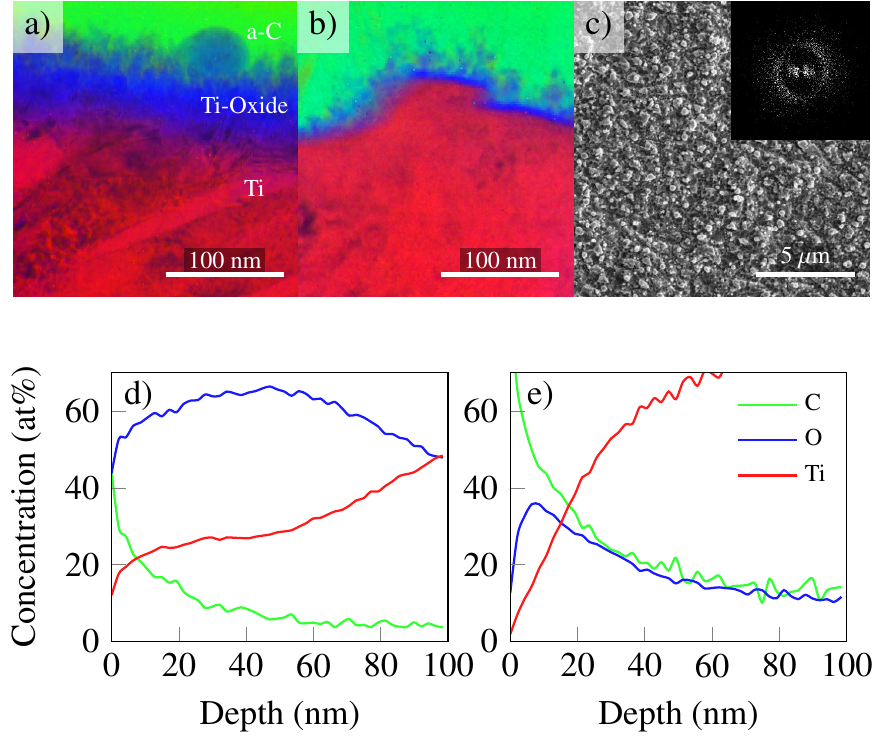}
	\vspace{-2mm}
	\caption{\label{fig:FIB_Oxi}EF-TEM images are shown for a bump in a) and for a reference sample with the same parameters but $v_\mathrm{s}$~=~3.5 m/s $\gg$ $v_{\mathrm{crit}}$ in b). Picture c) is an SEM image of the reference sample. An amorphous carbon layer was deposited prior to lamella preparation. Color coding: amorphous carbon (a-C) in green, titanium oxides (Ti-Oxide) in blue, and titanium (Ti) in red. The oxide-rich layer (blue) is much thicker for the bump. The graphs d) \& e) show the atomic concentrations of a) \& b) as a function of depth; they were determined by Auger electron spectroscopy. The lines represent an interpolation of 50 equidistant data points.}
	\vspace{-1mm}
\end{figure}

The EF-TEM images in Figure \ref{fig:FIB_Oxi} show a clear difference in oxide content (in blue) between laser-milled samples with bumps in a) and without bumps in b). Note that the layer thickness of oxides is < 200 nm in the case of bumpy surfaces and <~50~nm in the case of processed but smooth surfaces. At a depth of 100~nm, approximately 50\% oxygen is still present at  the bumpy surface. This high oxygen content is not reached in the smooth sample for any depth. Given that the intensity is the same in both cases and, thus, the energy penetration depth and the depth of fusion should be comparable, the increased oxygen concentration shown in Figure \ref{fig:FIB_Oxi} a) was necessarily affected by the scan speed and consequently by the heat accumulation. 

For a diffusion coefficient of oxygen in liquid Ti  ${D_{\mathrm{o}} \leq 10^{-5}\ \mathrm{cm^2/s}}$ \cite{zeng2020diffusion}, the diffusion depth is ${l_{\mathrm{diff}} = \sqrt{D_{\mathrm{o}} \tau_{\mathrm{s}}} \leq 10\ \mathrm{nm}}$
for a solidification time ${\tau_{\mathrm{s}} \leq 100\ \mathrm{ns}}$.
Consequently, the oxygen enriched layer 
is not formed solely by diffusion. Most likely, convection enhances the oxygen intermixing with the liquid titanium \cite{menold2018surface}.
The complex interplay of thermoelastic pressure waves \cite{rethfeld2017modelling,ivanov2003effect}, temperature gradients resulting from IED, thermocapillary waves \cite{levchenko1981instability,tsibidis2015ripples,tsibidis2016convection}, recoil, and evaporation pressure \cite{chen2009formation,gurevich2016mechanisms} can trigger a wide variety of convection processes. For a detailed discussion of convection processes in ultrashort pulse ablation, the interested reader is referred to Refs. \cite{rudenko2020high,bonse2020maxwell}. 

Convection processes, whose characteristic time can be approximated by $\tau_{\mathrm{M}}~\approx~\frac{\mu \Lambda^2}{4L\gamma\Delta T_{\mathrm{lat}}}$, with a temperature difference $\Delta T_{\mathrm{lat}}$ over a characteristic length $\Lambda$, dynamic viscosity $\mu$, surface tension coefficient $\gamma$, and melt depth $L$, could potentially determine both the groove formation and the thickness of the oxide layer. 

The interval required for the formation of instabilities on characteristic dimensions of the grooves ($\Lambda_{\mathrm{G}} \sim \lambda_{\mathrm{L}} \approx$~1~$\mu$m) and the oxide layer ($L_{\mathrm{oxid}} \leq$~200~nm) is within a temporal scale ranging from $\tau_{\mathrm{M}}$~=~1~to~50~ns for melt film depths of 50~-~500~nm. In contrast, the dimensions of the bumps themselves, $\Lambda_B$~$\gg$~1~$\mu$m, would result in characteristic times $\tau_{\mathrm{M}}$~$\gg$~100 ns. Consequently, the convective contributions to the bumps are more likely to be associated with the formation of grooves.

To correlate the time scales with the only varied parameter, i.e., the scan speed, an analytical model was applied to determine the melt film thickness and the solidification time for a single laser pulse impacting the surface in quasi-steady state ($T_{0} = T_{\mathrm{sat}}$) using an effective melting temperature $T_{\mathrm{eff}}~=~T_{\mathrm{m}}~+~\Delta H_{\mathrm{m}} / c_{\mathrm{p}} =~$~2690~K. A detailed explanation of the model can be found in \cite{bauer2015heat}, and further information is provided in the supplementary material. 

Figure \ref{fig:Meltdepth_and_Time} shows the solidification time and maximum melt depth. Apparently, the scan speed has a small impact on the melt depth, mainly influenced by the fluence \cite{anisimov1997theory} at moderate repetition rates, but a high influence on the solidification time. Consequently, the stronger heat accumulation at slow scan speeds delays solidification and thus provides more time for hydrodynamic instabilities to develop.

\begin{figure}[h!]
    \centering
	\includegraphics{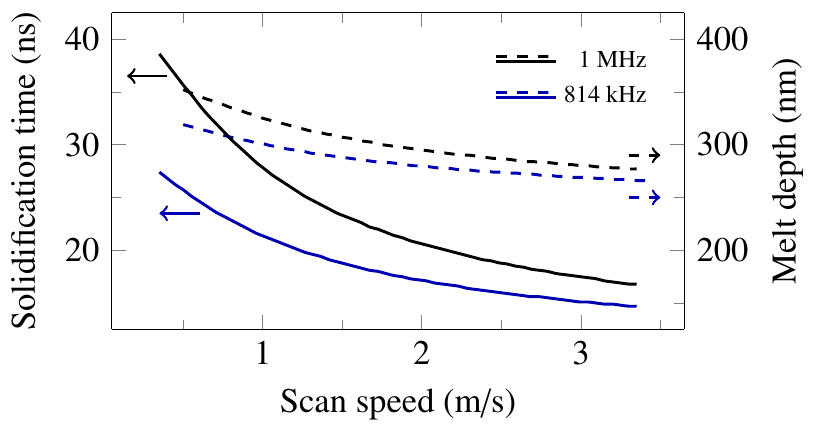}
	\vspace{-1mm}
	\caption{\label{fig:Meltdepth_and_Time} Solidification time (solid lines) and maximum melt depth (dashed lines) for different scan speeds; the laser parameters were the same as those set in Figure~\ref{fig:Titan_1MHz_zirkular_1mps}. The stronger heat accumulation at lower scan speeds indicates a much higher influence on the solidification time than on the melt depth. Averaged material properties were used for the calculations ($\rho$~=~4.4 g/cm$^3$, $D_{\mathrm{th}}$~=~8.22~mm$^2$/s, $c$~=~626~J/kg~K)~\cite{desai1987titanium}. }
\end{figure}

As mentioned in Sec. \ref{sec:Theo_view}, the formation process of grooves is mostly explained using hydrodynamic  models. Besides, the strong influence of the scan speed on their size shown in Figure \ref{fig:incr_vmin_ti} implies a connection with the heat accumulation and, according to Figure~\ref{fig:Meltdepth_and_Time}, with the solidification time. If we assume delayed solidification caused by heat accumulation as the reason for the development of hydrodynamic instabilities creating grooves of critical size, we can relate the critical spatio-temporal pulse overlap to a physical process. The critical size is to be understood as dimensions that lead to a further local increase of the fluence within the grooves, thereby causing the grooves to grow steadily due to non-uniform ablation. This critical size might be close to an aspect ratio $L_\mathrm{G}/\Lambda_{\mathrm{G}}$ $\geq$ 0.4 with $\Lambda_{\mathrm{G}}$ > $\lambda_{\mathrm{L}}$ as width, and $L_\mathrm{G}$ as depth of the grooves \cite{qin2012comparison}. However, further investigation is required to exactly determine this tipping point.  

For the same parameter values as those in Figure \ref{fig:Titan_1MHz_zirkular_1mps}, the critical spatio-temporal overlap requires a scan speed $v_\mathrm{s}$ $\geq$ 2.5~m/s for $f_{\mathrm{R}}$~=~1~MHz at $\omega_0$~=~20~$\mu$m. According to Figure \ref{fig:Meltdepth_and_Time}, this set of parameters leads to a solidification time $\tau_{\mathrm{s}}$~$\leq$~20~ns. Even if we assume a high temperature difference of $\Delta T_{\mathrm{lat}}$~$\approx$~10$^3$~K over $\Lambda$ $\geq$ 1 $\mathrm{\mu}$m and the remaining values as averaged values from the model, this leads to $L_{\mathrm{max}}$~=~220~nm, $\mu$~=~2.25~mPa$\cdot$s, and a characteristic time of $\tau_{\mathrm{M}}$~$\geq$~20~ns at $\gamma$~=~0.25~mN/Km. Consequently, high scan speed prevents the delay in solidification necessary for the development of instabilities. This hypothesis is also consistent with the observed dependence of bump formation on the initial temperature of the material \cite{bauer2015heat}.
Although this hypothesis explains the dependence of groove formation on spatio-temporal pulse overlap, the question of what drives convective mass transport remains unanswered. This is a question related to a potential electromagnetic origin of the grooves that is addressed in the next section.

\subsection{\label{ssec:Intensity-dependet-formation}Surface structure origin: IED-related structure formation}

Supposing an IED in the electron system (\textit{pri./sec. modulation} in Figure~\ref{fig:sketch}), strong electron-phonon coupling leads to strong energy confinement and thus to conservation of the IED during energy transfer to the lattice. This can be intuitively understood as beneficial for structure formation. Strong coupling was already discussed as a key parameter of strong LSFL formation \cite{colombier2012effects,wang2005ultrafast}. However, to date this has been modeled by assuming a constant coupling factor $G$  and through material comparisons. In our case, we assume a single material whose coupling factor depends on electron-temperature $T_\mathrm{e}$.  Therefore, we investigated whether it is possible to influence the structure formation by changing the pulse intensity and thus $T_\mathrm{e}$. For this purpose, we selected a parameter set characterized by a low repetition rate, $f_{\mathrm{R}}$~=~100~kHz, causing a lower heat accumulation and a solidification time that is primarily determined by the applied pulse energy \cite{anisimov1997theory,fang2017direct}. Contributions to structure formation from ablation pressure are also expected to increase with increasing fluence \cite{tsibidis2015ripples,tsibidis2016convection,tsibidis2018modelling}. Consequently, one would expect an earlier occurrence of bumps for higher fluences. 

\begin{figure}[b!]
\vspace{-12pt}
	\centering
	\includegraphics{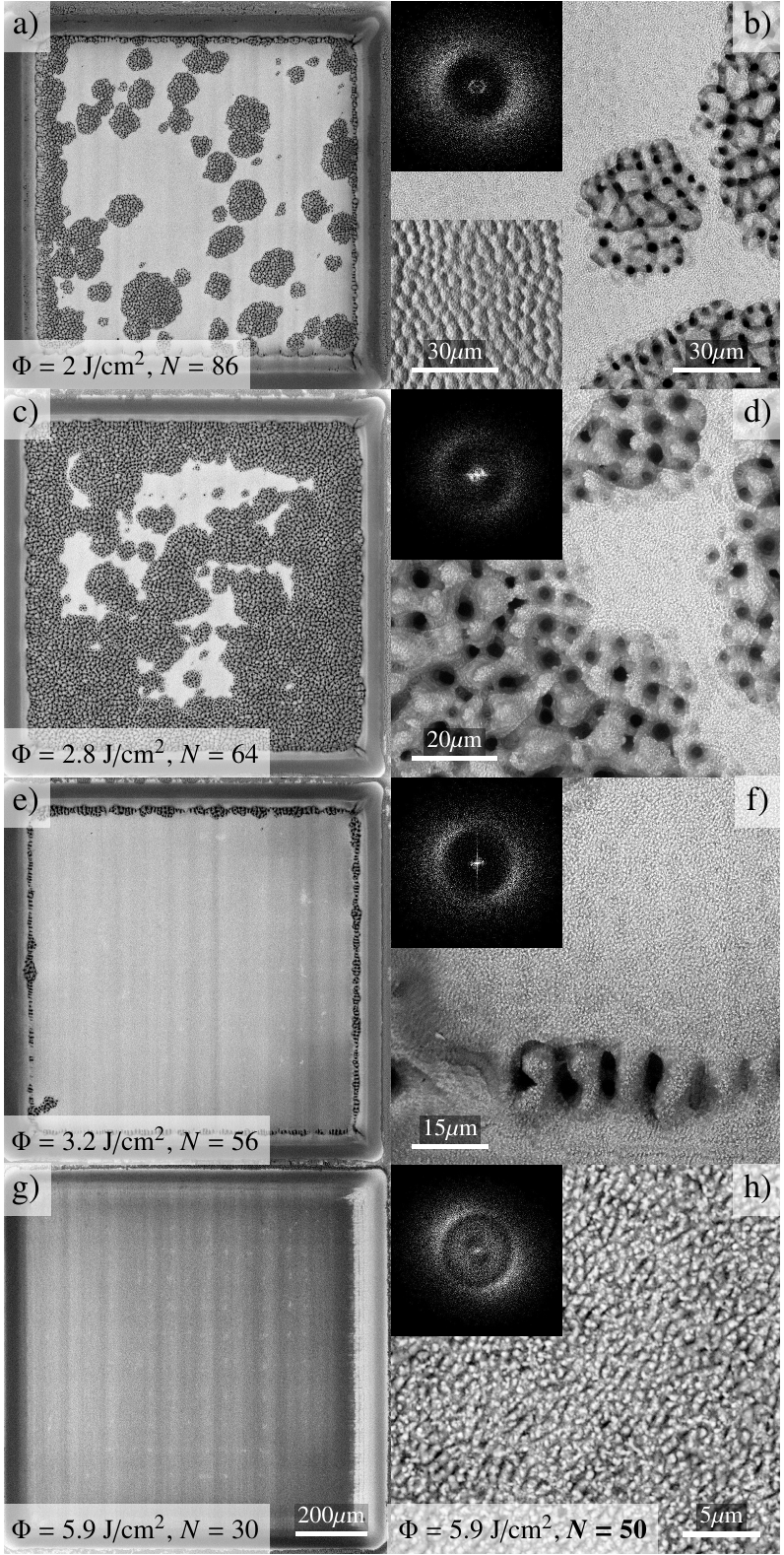}
	\caption{\label{fig:Ti_100kHz_milling} SEM images of milled squares on titanium with $\tau$~=~400~fs, $f_{\mathrm{R}}$~=~100~kHz, $v_\mathrm{s}$~=~0.5~m/s, $d_h$~=~5~$\mathrm{\mu}$m, 90$^{\circ}$-cross-hatch, and circular polarization. The number of scans $N$ was adjusted to the fluence $\Phi$ to keep the ablation depth comparable. A larger $N$ was used in h) to demonstrate that the bumps and grooves remain absent. In the right column, enlargements of representative areas are shown. The dark insets are FFTs of areas that are not covered with bumps. A similar area is also shown in image b) as an inset in topography mode (detector under a stronger angle, same magnification). It is remarkable that the bumps disappear again for very high fluences or intensities. The LSFL pattern in h) indicates that the IED within the ablation depth has not disappeared. The FFT of picture h) does not show any grooves features (type-g), as it is the case in the FFTs of b) \& d).}
\end{figure}
Figure \ref{fig:Ti_100kHz_milling} shows SEM images of milled Ti for increasing pulse energies. 
The left column in Figure \ref{fig:Ti_100kHz_milling} shows the entire ablated area. The dark areas exhibit bumps which, in contrast to the observations for steel \cite{fraggelakis2017texturing,ackerl2020evolution}, decrease with increasing fluence. Possible reasons for that anomaly of Ti are discussed below. 

Representative areas are shown as enlargement in the second column. FFTs, shown as insets, were obtained over areas not covered with bumps yet. In Figure \ref{fig:Ti_100kHz_milling} a), the bumps appear in groups. These groups merged together, leaving only a few areas without bumps, as depicted in Figure \ref{fig:Ti_100kHz_milling} c). Given that the LSFL formation depends on crystal orientation and distance to grain boundaries, this group-wise occurrence is expected for poly-crystalline materials~\cite{sedao2014influence}. Recently, the influence of line-like defects and grain distribution on the formation of LSFL and larger structures was studied and discussed extensively~\cite{ackerl2020observation}.

The lower inset in Figure \ref{fig:Ti_100kHz_milling} b) shows a recording under a higher detector angle (same magnification), making the grooves much more visible. Note also that the type-g signals are again strongly represented in the FFTs (signal near the center), especially in Figure~\ref{fig:Ti_100kHz_milling}~d). It seems that the bumps grew out of the grooves again. Although the heat accumulation was lower at 100~kHz, the higher fluences led to a strong delay in solidification \cite{fang2017direct}. This could be the reason for the high fluence threshold for grooves usually reported in previous studies for low repetition rates \cite{bonse2005structure,nivas2021generation}.

In Figure \ref{fig:Ti_100kHz_milling} e) and h), only the edges of the ablated square exhibit large structures. This effect is sometimes referred to as edge deepening and is associated with stronger polarized reflections from the walls \cite{nyenhuisdual}. 

Concerning Figure \ref{fig:Ti_100kHz_milling} h), 50 scans were applied to verify that the bumps were absent even for a high number of scans. The surface shown in Figure \ref{fig:Ti_100kHz_milling} h) exhibits no larger structures. There are no signs of the presence of type-g signals or grooves. Both the FFT and the surface show only slight features of LSFL, which become clear with respect to a uniformly distributed residual signal. 

The results of these experiments seem counterintuitive. Usually, for fluences above the optimum fluence, the residual heat increases \cite{bauer2015heat,neuenschwander2012optimization}, thereby causing a strong delay in solidification \cite{fang2017direct}. Consequently, one would expect more bumps with increasing fluences in Figure \ref{fig:Ti_100kHz_milling}, as is normally the case \cite{fraggelakis2017texturing,ackerl2020evolution}. Although fluences are compared in the figure, the peak intensity also changed as the pulse energy was successively increased to raise the fluence. These experimental results indicate a change either in the IED or in the electron-phonon coupling with rising fluence or intensity, respectively. To support this interpretation of the experiments, we make use of electron-phonon coupling modeling.

\subsection{Theoretical modeling}

We investigated the reduction of lateral temperature gradients from IED due to electron diffusion by simulating Ti heating following ultrafast laser pulse irradiation in the framework of the two-temperature model (TTM) \cite{anisimov1974electron,rethfeld2017modelling}:
\begin{equation}
    C_\mathrm{e} \frac{\partial T_\mathrm{e}}{\partial t} = \nabla \cdot ( \kappa_\mathrm{e} \nabla T_\mathrm{e}) - G(T_\mathrm{e} - T_{\mathrm{lat}}) + S,\label{eq:TTM_E}
\end{equation}
\begin{equation}
\label{eq:TTM_L}
    C_{\mathrm{lat}} \frac{\partial T_{\mathrm{lat}}}{\partial t} = \nabla \cdot ( \kappa_{\mathrm{lat}} \nabla T_{\mathrm{lat}}) + G(T_\mathrm{e} - T_{{\mathrm{lat}}}),
\end{equation}
where $T_\mathrm{e}$ and $T_{\mathrm{lat}}$ are the electron and lattice temperatures; $C_e$ and $C_{\mathrm{lat}}$ are the electron and lattice heat
capacities; $\kappa_e$ and $\kappa_{\mathrm{lat}}$ are the electron and lattice heat conductivities, respectively; and $S$ is the source term, which describes laser energy absorption. We assume a cosine-like function in the lateral {$x$-direction} to study relaxation dynamics of laser-induced periodic energy deposition (further details are provided in the supplementary material). 

The temporal pulse spacing $\Delta t$~=~10~$\mu$s at 100 kHz allows the most important conclusions about the process to be drawn from the observation of a single pulse, given that shielding effects or strong heat accumulation take place at much shorter $\Delta t$ \cite{forster2018shielding,bauer2015heat}.

\begin{figure}[b]
	\centering
	\includegraphics{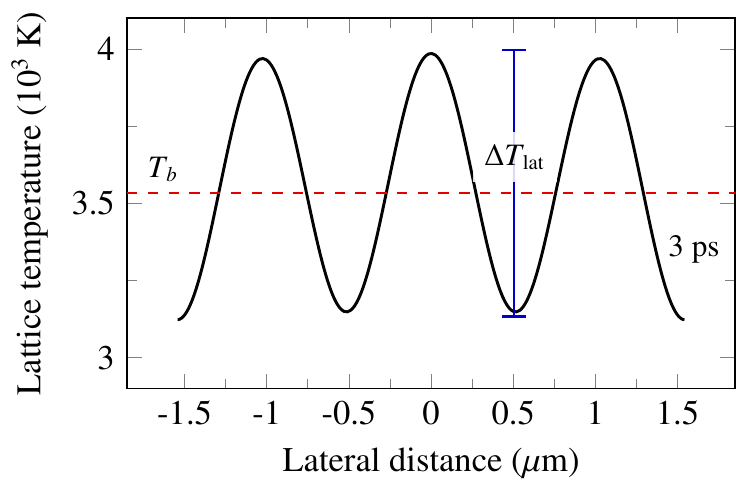}
	\caption{Lattice temperature profile at 
	$z$~=~7.5~nm and at 3 ps. Laser parameters: wavelength of 1030~nm, pulse duration of 400~fs, and fluence of 0.08~J/cm$^2$; $\Delta T_{\mathrm{lat}}$ denotes the amplitude of the lattice temperature oscillations.\label{fig:sim1}}
\end{figure}

Figure \ref{fig:sim1} represents the profile of the lattice temperature along the lateral $x$-direction at a depth of $z$~=~7.5~nm and at 3~ps after applying a single laser pulse with a wavelength of 1030~nm, a duration of 400~fs, and a fluence of 0.08~J/cm$^2$. We can observe in Figure \ref{fig:sim1} oscillations of $T_{\mathrm{lat}}$ around the boiling temperature $T_\mathrm{b}$ = 3560~K for Ti.

\begin{figure}[h!]
	\centering
	\includegraphics{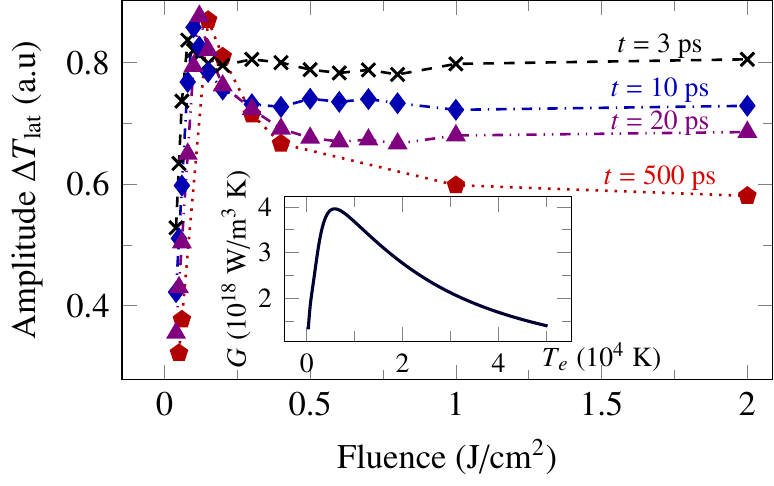}
	\caption{Lattice temperature amplitude $\Delta T_{\mathrm{lat}}$ as a function of incident laser fluence at different time instants, denoted as $t$, after absorption. The data points were connected by a dashed line to drive the reader's eye. A sharp decay of the amplitude with increasing fluence can be observed. The electron-phonon coupling factor $G$ as a function of electron temperature is shown as an inset, following Lin $et.$ $al.$’s work \cite{lin2008electron}. \label{fig:sim2}}
\end{figure}

Figure \ref{fig:sim2} depicts the amplitude of the lattice temperature oscillations $\Delta T_{\mathrm{lat}}$ around the boiling temperature as a function of fluence. At low fluences, when $T_{\mathrm{lat}}$ does not reach the boiling point, the amplitude $\Delta T_{\mathrm{lat}}$ was taken at the sample surface.

Note that the amplitude of oscillations is decreasing with increasing laser fluence after reaching a peak value. This trend is in accordance with experimental observations and shows that the sample surface becomes smoother for high laser fluences. The origin of this trend is based on the dependence of the coupling factor $G$ on the electron temperature, which is shown in Figure~\ref{fig:sim2} as an inset. This dependence suggests that at high laser fluences the coupling factor decreases, which results in slow electron-lattice energy transfer. Therefore, the hot electron ensemble has more time for diffusion and releases its energy over a larger affected sample volume. As a result, we observe the decreasing trend of the lattice temperature amplitude, which might explain the reduced structure formation.

We did not consider phase transitions in the applied model. Therefore, artificial heating of deeper layers by non-removed (ablated) hot upper layers might emerge with increasing simulation time. Given that the electron-phonon equilibrium time and the depth of $\Delta T_{\mathrm{lat}}$ increase with fluence, and that electron diffusion over several nanometers requires a few picoseconds \cite{ivanov2003combined,brorson1987femtosecond}, we also need to consider solutions at later time instants. However, the major decrease in $\Delta T_{\mathrm{lat}}$ occurs within the first 10~ps. Simulations \cite{schaefer2002metal,ivanov2003combined} and experiments \cite{sokolowski1998transient,winter2020ultrafast} indicate that the ablated material loses its connection with the bulk for $t$~>~10~ps after absorption. 

More detailed simulations should also account for a change in optical properties with increasing intensity, which would affect the initial amplitude of the modulation.

Ti, as a transition metal, exhibits d-bands with a high density of states near the Fermi level. This allows inter-band transitions between the highly populated d-band $g_d~\approx~2$~eV$^{-1}$ to the less populated s-band $g_s \approx$~0.5~eV$^{-1}$ \cite{aguayo2002elastic,lin2008electron}. 
As a result, $\tilde{\epsilon}$ changes strongly with the laser pulse intensity \cite{gnilitskyi2017high,golosov2011ultrafast,bevillon2018nonequilibrium}, which affects the SPP excitation and thus most likely the formation of LSFL \cite{lalanne2009microscopic,golosov2011ultrafast,derrien2016properties}. Less or lower pronounced LSFL would also affect the grooves formation, given that deep LSFL ($\geq$~100~nm) are necessary for the occurrence of type-g signals in FDTD simulations \cite{zhang2015coherence}. In addition, it is well known that LSFL occurs only in a relatively narrow intensity range; this has been attributed to the change in $\tilde{\epsilon}$ \cite{bonse2009role}. It was also experimentally demonstrated that the LSFL regularity depends on the free mean path length of SPP \cite{gnilitskyi2017high}. All this indicates that SPP plays an important role in positive feedback mechanisms of LSFL formation. 

Recently, a strong dependence of the optical properties of transition metals on $T_\mathrm{e}$ was investigated through \textit{ab initio} molecular dynamic simulations coupled to density functional theory and Kubo-Greenwood formalism \cite{bevillon2018nonequilibrium,bevillon2016ultrafast}. As a general trend, a decreasing influence of the inter-band effects on the optical properties was identified with increasing $T_\mathrm{e}$. 

\begin{figure}[h!]
    \centering
    \includegraphics{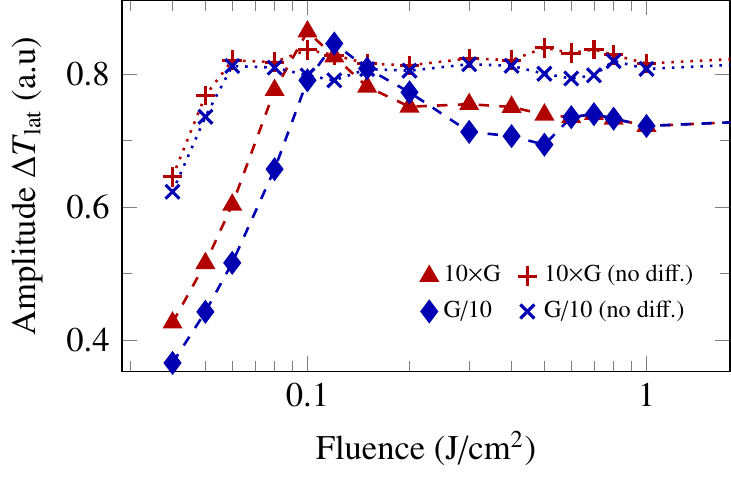}
    \caption{Lattice temperature amplitude $\Delta T_{\mathrm{lat}}$ as a function of incident laser fluence at $t$~=~10~ps for $G$/10 and 10$\times G$ with and without the influence of diffusion ($\kappa_{\mathrm{lat}}=\kappa_\mathrm{e}=0$). The data points were connected by dashed and dotted lines for emphasis. }
    \label{fig:sim3}
\end{figure}

A high $T_\mathrm{e}$ is also associated with a strong change in $G$ \cite{lin2008electron}. Therefore, it is experimentally difficult to separate the influence of $G$ and $\tilde{\epsilon}$ on LSFL formation. Thus, Figure \ref{fig:sim3} provides a numerical view of the influence of $G$ based on its variation over two orders of magnitude considering and neglecting the electron and lattice heat conductivities, or, in other words, a diffusion in the electron and lattice subsystems. 
For fluences below the ablation threshold of 0.1 J/cm$^2$ \cite{nyenhuis2021ultrashort}, we observe that $G$ plays an important role resulting in higher values of $\Delta T_{\mathrm{lat}}$ for higher values of $G$. In addition, it can be seen that an artificially neglected diffusion ($\kappa_{\mathrm{lat}}=\kappa_\mathrm{e}=0$) leads to increased values of $\Delta T_{\mathrm{lat}}$ compared to the case with a diffusion. In the fluence range above the ablation threshold, approximately, from 0.1 J/cm$^2$ till 0.5 J/cm$^2$, the values of $\Delta T_{\mathrm{lat}}$ are close to each other for all considered cases. This indicates a complex interplay of influences of $G$ and diffusion. However, in general, higher value of $G$ result in higher $\Delta T_{\mathrm{lat}}$. \newline
\indent In the high fluence range (above 0.6 J/cm$^2$), the values of  $\Delta T_{\mathrm{lat}}$ are almost the same for the $10\times G$ and $G$/10 cases with diffusion. Therefore, we can conclude that the importance of $G$ reduces to a minimum in the high fluence range. In contrast, $\Delta T_{\mathrm{lat}}$ are slightly different for the $10\times G$ and G/10 cases without a diffusion. Nevertheless, we observe experimentally a significant change in surface morphology within the high fluence range. This points out to a non-negligible influence of the change in $\tilde{\epsilon}$ during femtosecond photoexcitation at higher intensities.

\section{Summary}

In this contribution, a model of bump formation and the associated empirical critical saturation temperature $T_{\mathrm{c}}$ in usp milling is provided. Using a hierarchical structural model, the individual steps and the underlying physical mechanisms are discussed. A delay in solidification due to strong heat accumulation enabling convection mechanisms is identified as the reason for growth of grooves, which act as transition structures from LIPSS to bumps. The non-perpendicular orientation between grooves and LSFL in our experiments may be considered as another hint for an electromagnetic origin of the grooves \cite{zhang2020laser}.

Furthermore, it is shown that the whole structure formation is sensitive to the applied intensity. Using a TTM with modulated source, the smearing of inhomogeneous energy deposition by electron diffusion is discussed. Our simulation results indicate a moderate influence of the coupling behavior. This once again highlights the apparently important role of SPP in the positive feedback mechanism of LSFL evolution, given that we consequently assume a strong change in optical properties as the cause of the reduced structure formation.

Our findings shed new light on the choice of laser wavelength for usp milling processes, because the excitation efficiency of SPP scales with $\tilde{\epsilon}$ \cite{lalanne2009microscopic}, the evolution of which in turn depends on the degree of excitation. This motivates further research on $\tilde{\epsilon}$ for femtosecond photoexcited transition metals and industrial relevant alloys. 

In addition, the intensity-related structure formation experiments stress the need for LSFL as an initial nucleation structure for surface degradation. In this way, it was concluded that the surface degradation is not only of thermal origin. Therefore, a dual process strategy of polishing and ablation steps is suitable for further increase of the ablation rates~\cite{nyenhuisdual}, given that, in the absence of LSFL, the residual heat of previous polishing steps does not trigger bump formation.

\section*{Acknowledgments}

We acknowledge the financial support of the Deutsche Forschungsgemeinschaft project RE1141/14-2. Simulations were performed on the high-performance cluster ``Elwetritsch'' through projects TopNano and Mulan at the TU Kaiserslautern, which is part of the ``Alliance of High Performance Computing Rheinland-Pfalz''. We kindly acknowledge the support of Regionales Hochschulrechenzentrum Kaiserslautern.

P.N.T. would like to thank Vladimir Lipp for discussions regarding the implementation of a numerical scheme to find a solution of 2D TTM.

\section*{Conflict of Interest}
The authors declare no conflict of interest.

\bibliography{References2}
\clearpage



\section*{Supplementary material}

# Fundamentals of Scanning Surface Structuring by Ultrashort Laser Pulses: From Electron Diffusion to Final Morphology

**F. Nyenhuis**[*]
Robert Bosch GmbH, Bosch Research,
Postbox 30 02 40, Stuttgart, 70442, Germany

**P. N. Terekhin**
Department of Physics and Research Center OPTIMAS,
Technische Universität Kaiserslautern,
Erwin-Schrödinger-Str. 46, Kaiserslautern, 67663, Germany

**T. Menold**
Robert Bosch GmbH, Bosch Research,
Postbox 30 02 40, Stuttgart, 70442, Germany

**B. Rethfeld**
Department of Physics and Research Center OPTIMAS,
Technische Universität Kaiserslautern,
Erwin-Schrödinger-Str. 46, Kaiserslautern, 67663, Germany

**A. Michalowski**
Robert Bosch GmbH, Bosch Research,
Postbox 30 02 40, Stuttgart, 70442, Germany

**J. L'huillier**
Photonik-Zentrum Kaiserslautern e.V. and Research
Center OPTIMAS, Technische Universität Kaiserslautern,
Kohlenhof Strasse 10, Kaiserslautern, 67633, Germany

January 18, 2022

## Supplemental Materials

### Experimental Setup

All experiments were performed at a central wavelength of $\lambda_L$ = 1030 nm with a $M^2 \leq 1.3$ at ambient atmosphere. A pulseCheck (APE, Angewandte Physik & Elektronik GmbH) was used to measure the pulse duration. Different pulse duration measurements exhibit a maximum variance of 10% for $\tau \geq 1$ ps and 15% $\tau <$ 1 ps. An intelliSCAN 14 scanner system (SCANLAB GmbH) and a F-Theta lens with a focal length of 163 mm were used for beam guidance. This leads to a spot size $2\omega_0$ of 40 $\mu$m (1/e$^2$ diameter) and a Rayleigh length $\omega_R$ of about 1 mm. The spot size was checked several times using Liu's method [S1]. A laser scanning microscope (LEXT OLS4000, Olympus) was applied for evaluation of the Liu method (using a MPlan Apo 100, Na = 0.95).

### Sample preparation & analytical equipment

Titanium sheets with a purity of over 99.95% and a thickness of 1 mm were used. The samples were mechanically polished to an arithmetical mean roughness value of $R_a \leq 60$ nm for evaluation of structure formation (Fig. 2 - 4, in the main text).

For the analysis of the bumps, a focused ion beam (FEI 875 dual-beam Focused Ion Beam, Ga+ ions) was used to cut about 20 $\mu$m wide and approx. 5 -7 $\mu$m deep mirco-cross sections. Also the FIB was used to prepare lamellae. Prior to FIB preparation, a layer of amorphous carbon was deposited on the sample surface (1st protective layer). To avoid ion-induced damage to the near-surface regions of a lamella, a protective tungsten layer was deposited locally by the focused ion beam before the actual lamella preparation (2nd protective layer). The transmission electron images

---

[*]fabian.nyenhuis@gmail.com

(U = 197 keV, collection half angle -13 mrad), as well as the energy filtered images, were taken with a Joel 2010 in analytical version Gatan imaging filter (GIF-Tridiem 863) with a slow-scan CCD camera (Model 794) from Gatan (Oxford Instruments). Auger electron spectroscopy (AES) has been used for lateral and depth-resolved characterization of laser-treated titanium surfaces. For this purpose, a combination of SE images, AES point and ROI spectra, and sputter ablation is used for depth profiling. The information depth of AES is a few nm and the lateral resolution of the method is about 30 nm. The detection limit is about 0.1-0.5 at%.

## Data evaluation

The discrete Fourier transforms were determined from images at 2000 and 3000 magnification (23 and 15.3 pixels per micrometer) of the SEM images (Phenom) using ImageJ [S2]. The dispersion LIPSS orientation angles $\delta\Theta$ (DLOA) were determined using the ImageJ OrientationJ plugin [S3] in the same way as by Gnilitskyi *et al.* [S4]. For the evaluation 53×53 $\mu$m SEM images were used at a 5000 magnification. In order to measure only the angle of the LSFL without the influence of the grooves, a high-pass filter was previously applied ($\lambda_{cutoff} = 1~\mu m$). The further procedure is analogous to Gnilitskyi *et al.* [S4]. The analytical model as described in [S5] was used to determine the surface temperature. In all cases, a residual heat of $E_{res} = (1-R) \cdot E_p = 0.39 \cdot E_p$ was assumed as an upper estimate. The used material properties are given in the figure captions.

## Details of two-temperature simulations

We describe the laser energy absorption upon irradiation of Ti sample by applying the source term with a Gaussian shape:

$$S(x,z,t) = \Phi_{\text{inc}}\alpha_{\text{inc}}(1-R)\Psi(x)e^{\alpha_{\text{abs}}z}e^{\frac{-2x^2}{\omega_0^2}}\frac{1}{\tau}\sqrt{\frac{\sigma}{\pi}}e^{\frac{-\sigma(t-t_0)^2}{\tau^2}}, \tag{S1}$$

where $\Phi_{\text{inc}}$ is the incident peak laser fluence

$$\Phi_{\text{inc}} = \frac{2E_p}{\pi\omega_0^2} \tag{S2}$$

with $E_p$ being the total pulse energy and $\omega_0$ being the 1/e laser field radius. The value $\alpha_{\text{abs}} = 2k_0 k_m$ is the absorption coefficient, $k_0 = \omega/c$ is the wave vector of light, $\omega$ is the laser angular frequency, $c$ is the speed of light, $n_m$ and $k_m$ are the real and imaginary parts of the complex refractive index of Ti $\tilde{n}_m = n_m + ik_m$. We define the reflectivity R as

$$R = \frac{(n_m - 1)^2 + k_m^2}{(n_m + 1)^2 + k_m^2}. \tag{S3}$$

The function $\Psi$ is responsible for the assumed laser-induced periodic energy deposition in the lateral $x$-direction:

$$\Psi(x) = 1 + \eta_e \cos\left(\frac{2\pi}{\lambda_x}x\right), \tag{S4}$$

where the parameter $\eta_e$ describes the amplitude of the source term modulation along the $x$-axis and $\lambda_x$ is the periodicity of modulation. The value $\tau$ stands for a pulse duration at FWHM (full width at half maximum), $t_0 = 3\tau$ chosen as a location of the pulse maximum and $\sigma = 4\ln 2$. All the parameters used in our simulations are presented in Table S1.

We have calculated the periodicity of modulation $\lambda_x$ as

$$\lambda_x = \frac{2\pi}{k'_x}, \tag{S5}$$

where $k'_x$ is the real part of the surface plasmon polariton (SPP) wave vector

$$k_x = k_0\sqrt{\frac{\epsilon_m\epsilon_d}{\epsilon_m + \epsilon_d}} \tag{S6}$$

with $\epsilon_m$ being the dielectric function of Ti [S8] and $\epsilon_d$ being the dielectric function of the dielectric (for air we set $\epsilon_d = 1$).

The TTM equations 1 and 2 from the main text were solved numerically assuming that initially electron and lattice subsystems were in the equilibrium at 300 K. No-flux boundary conditions were applied:

Table 1: Simulation parameters for Ti at 1030 nm laser wavelength and 400 fs pulse duration.

| Quantity | Symbol | Value | Ref. |
|---|---|---|---|
| Electron heat capacity | $C_e$ | Taken from Ref. [S6] | [S6] |
| Lattice heat capacity | $C_{\text{lat}}$ | $2.35 \times 10^6$ J/(m$^3$K) | [S7] |
| Electron heat conductivity | $\kappa_e$ | Taken from Ref. [S7] | [S7] |
| Lattice heat conductivity | $\kappa_{\text{lat}}$ | 22 W/(m K) | [S7] |
| Electron-lattice coupling factor | $G$ | Taken from Ref. [S6] | [S6] |
| Absorption coefficient | $\alpha_{\text{abs}}$ | $4.86 \times 10^7$ m$^{-1}$ | [S8] |
| Reflectivity | $R$ | 0.61 | [S8] |
| Modulation amplitude | $\eta_e$ | 0.15 | [S7] |
| Periodicity of modulation | $\lambda_x$ | 1028 nm | [S8] |
| 1/e laser field radius | $\omega_0$ | 20 $\mu$m | Our work |

$$\frac{\partial T_e(x_{\min}, z, t)}{\partial x} = \frac{\partial T_e(x_{\max}, z, t)}{\partial x} = \frac{\partial T_e(x, z_{\min}, t)}{\partial z} = \frac{\partial T_e(x, z_{\max}, t)}{\partial z}, \quad (S7)$$

$$\frac{\partial T_{\text{lat}}(x_{\min}, z, t)}{\partial x} = \frac{\partial T_{\text{lat}}(x_{\max}, z, t)}{\partial x} = \frac{\partial T_{\text{lat}}(x, z_{\min}, t)}{\partial z} = \frac{\partial T_{\text{lat}}(x, z_{\max}, t)}{\partial z}, \quad (S8)$$

where $x_{\min}$ = -1542 nm and $x_{\max}$ = 1542 nm are the minimum and maximum coordinates along the $x$-axis, $z_{\min}$ = 0 nm and $z_{\max}$ = - 400 nm are the minimum and maximum coordinates along the $z$-axis, which is directed outside the sample. We have chosen the value $z_{\max}$ big enough that both electron and lattice thermal diffusion cannot reach this depth during our simulations. TTM equations were solved using the alternating-direction implicit method [S9]. It should be noted that our situations do not take into account phase transitions (melting and boiling). For the sake of simplicity, we assumed the constant optical properties (dielectric function, reflectivity and absorption coefficient). In addition, if the electron temperature reaches the maximum values given in Ref. [S6] for the electron heat capacity $C_e$ and the electron-lattice coupling factor $G$, we take these values at $T_{e,\max}$ for higher electron temperatures. Nevertheless, our simulations are able to describe the experimentally observed trend of the forming structures on the surface after ultrafast laser irradiation.


\end{document}